\documentclass[twocolumn]{aastex63}
\usepackage{amsmath}
\usepackage{stix}
\usepackage[symbol]{footmisc}
\usepackage{hyperref}

\usepackage{etoolbox}
\AtBeginEnvironment{thebibliography}{\linespread{1}\selectfont}
\received{1 July 2021}
\revised{6 October 2021}
\accepted{14 October 2021}
\submitjournal{ApJ}

\shorttitle{VLA and NOEMA view of the Bok Globule CB 17}
\shortauthors{Spear et al.}
\graphicspath{{./}{figures/}}

\begin{document}
\renewcommand{\thefootnote}{\fnsymbol{footnote}}
\title{VLA and NOEMA view of the Bok Globule CB 17: the starless nature of a proposed FHSC candidate \footnote{This work is based on observations carried out under project number S20AD with the IRAM NOEMA Interferometer. IRAM is supported by INSU/CNRS (France), MPG (Germany) and IGN (Spain).}}
\renewcommand{\thefootnote}{\arabic{footnote}}

\correspondingauthor{Stephanie Spear}
\email{stephanie.spear@aya.yale.edu}

\author{Stephanie Spear}
\affiliation{Yale University Department of Astronomy, New Haven, CT 06511, USA}

\author{Mar\'ia Jos\'e Maureira}
\affiliation{Max-Planck-Institut f\"ur extraterrestrische Physik, Giessenbachstrasse 1, 85748 Garching, Germany}

\author{H\'ector G. Arce}
\affiliation{Yale University Department of Astronomy, New Haven, CT 06511, USA}

\author{Jaime E. Pineda}
\affiliation{Max-Planck-Institut f\"ur extraterrestrische Physik, Giessenbachstrasse 1, 85748 Garching, Germany}

\author{Michael Dunham}
\affiliation{Center for Astrophysics | Harvard \& Smithsonian, Cambridge, MA 02138, USA}
\affiliation{Department of Physics, State University of New York Fredonia, Fredonia, New York 14063, USA}

\author{Paola Caselli}
\affiliation{Max-Planck-Institut f\"ur extraterrestrische Physik, Giessenbachstrasse 1, 85748 Garching, Germany}

\author{Dominique Segura-Cox}
\affiliation{Max-Planck-Institut f\"ur extraterrestrische Physik, Giessenbachstrasse 1, 85748 Garching, Germany}



\begin{abstract}

We use 3mm continuum NOEMA and NH$_3$ VLA observations towards the First Hydrostatic Core (FHSC) candidate CB 17 MMS to reveal the dust structure and gas properties down to 600-1,100 au scales and constrain its evolutionary stage. We do not detect any compact source at the previously identified 1.3 mm point source, despite expecting a minimum signal-to-noise of 9. The gas traced by NH$_3$ exhibits subsonic motions, with an average temperature of 10.4 K. A fit of the radial column density profile derived from the ammonia emission finds a flat inner region of radius $\sim$1,800 au and a central density of $\sim$6$\times10^5$ cm$^{-3}$. Virial and density structure analysis reveals the core is marginally bound ($\alpha_{vir}$= 0.73). The region is entirely consistent with a young starless core, hence ruling out CB 17 MMS as a FHSC candidate. Additionally, the core exhibits a velocity gradient aligned with the major axis, showing an arc-like structure in the p-v diagram and an off-center region with high-velocity dispersion caused by two distinct velocity peaks. These features could be due to interaction with the nearby outflow, which appears to deflect due to the dense gas near the NH$_3$ column density peak. We investigate the specific angular momentum profile of the starless core, finding that it aligns closely with previous studies of such radial profiles in Class 0 sources. This similarity to more evolved objects suggests that motions at 1,000 au scales are determined by large-scale dense cloud motions and may be preserved through the early stages of star formation.

\end{abstract}


\section{Introduction}\label{sec:intro}
Low-mass star formation begins with the fragmentation and gravitational collapse of dense, unstable molecular cloud regions into cores (\citealt{Larson1969NumericalProto-Star}; \citealt{Adams1987SpectralObjects}; \citealt{Shu1987StarTheory};  \citealt{Bergin2007ColdFormation}; \citealt{Caselli2008SurveyCores}). As these starless cores undergo gravitational collapse, the central region increases in density. Once the density of the central region reaches about 10\textsuperscript{-13} g cm\textsuperscript{-3} it is opaque enough that the heat generated by the collapse can no longer be freely radiated away. This decreases cooling efficiency, which subsequently results in a rapid increase in gas temperature and pressure in the center. Eventually the temperature and pressure are sufficient enough to stop the collapse of the central region, forming a core in quasi-hydrostatic equilibrium known as the first hydrostatic core or FHSC (\citealt{Larson1969NumericalProto-Star}; \citealt{Masunaga1998ACollapse}; \citealt{Saigo2006EvolutionCores}; \citealt{Tomida2010RADIATIONOUTFLOW}). During this short-lived stage of quasi-hydrostatic equilibrium the core continues to accrete mass from the surrounding envelope. Once the central temperature of the core reaches around 2000 K the core material becomes unstable and begins to collapse again, channelling most of the energy from this process into the dissociation of H\textsubscript{2}. This second collapse of the smaller, internal region of the first core forms a second hydrostatic core known as a protostar (\citealt{Larson1969NumericalProto-Star}).

The first hydrostatic core (hereafter FHSC) is thus theorized as an intermediary stage of quasi-hydrostatic equilibrium between the prestellar and protostellar phases of low-mass star formation. Theoretical studies and numerical simulations demonstrate that FHSCs should have several defining characteristics, including: 1) a lifetime of about 10\textsuperscript{3} to 10\textsuperscript{4} years; 2) a radius of $\gtrsim 10 - 20$au; 
3) a small mass of 0.01 - 0.1 $M_{\odot}$; 
and 4) a low internal luminosity of $10^{-4} - 10^{-1} L_{\odot}$ (\citealt{Masunaga1998ACollapse}; \citealt{Machida2008HighCloud}; \citealt{Sakai2010DistributionsL1527}; \citealt{Pezzuto2012HerschelCloud};
\citealt{Dunham2014TheHerschel}; \citealt{Young2019SyntheticCalculations}). Numerical simulations also show that a slow (v\textsubscript{out}$ \sim$3-5 kms\textsuperscript{-1}), compact outflow may be launched during this stage. Additionally, a disk may be present during this FHSC stage or may form directly after, out of the FHSC itself (\citealt{Machida2008HighCloud}; \citealt{Machida2011TheDisc}; \citealt{Joos2012ProtostellarCollapse}). It follows that the properties of the  dense gas surrounding FHSCs should be between those of prestellar and protostellar cores.

While the existence of the FHSC was predicted over 40 years ago \citep{Larson1969NumericalProto-Star} a bona fide FHSC has yet to be observationally confirmed. Their faintness and short lifespans have been limiting in this respect, as we expect to see only one FHSC for every 18 to 540 protostars \citep{Enoch2010ACore}. Studies have indentified 13 FHSC candidates \citep{2006BellocheEvolutionary,Enoch2010ACore,Chen2010L1448Core,Pineda2011THEVeLLO,Pezzuto2012HerschelCloud,Chen2012SUBMILLIMETERCORE,Young2018WhatNature,2018AFriesenALMA,2020FujishiroLowVelocity}. Each of these candidates had spectral Energy Distributions (SEDs) consistent with the predictions for the FHSC stage. Accordingly, all of these sources have low intrinsic luminosities (\textless 0.1 \(\textup{L}_\odot\)). Additionally, clear outflows have been detected for 6 of these sources \citep{Dunham2011DetectionCore,Pineda2011THEVeLLO,Busch2020TheCha-MMS1,2015GerinNascentBipolar,2020FujishiroLowVelocity}. Interferometric observations using high-density tracers showed that some of these sources were indeed in a young evolutionary stage, either FHSC or young Class 0 protostar \citep{Pineda2011THEVeLLO,Chen2012SUBMILLIMETERCORE,2012SchneeStarless,2013HuangProbing,Vaisala2014High-resolutionChamaeleon-MMS1,Maureira2017KinematicsL1451-mm,JoseMaureira2017A58}. However, follow-up studies at higher resolution provided evidence for ruling-out most of the candidates with outflow detections. These follow-up studies have focused on higher resolution observations of the molecular outflows (if present) and the physical properties of the inner envelope gas and dust emission as a means of ruling out several FHSC candidates (\citealt{Pineda2011THEVeLLO};\citealt{Marcelino2018ALMAB1b-S}; \citealt{Hirano2019TwoALMA}; \citealt{Busch2020TheCha-MMS1}; \citealt{Allen2020HighChamaeleon-MMS1}; \citealt{Maureira2020ALMACandidates};\citealt{2020FujishiroLowVelocity}). Some of these studies show that the detected outflows are too fast to be consistent with the theoretical predictions for the FHSC. Others show that the molecular line emission and distribution of the source is in better agreement with a slightly more advanced evolutionary stage of a young Class 0 star. Additionally, follow-up studies for two of these sources  lacked detection of a compact dust structure, thus demonstrating that these sources are consistent with starless condensations as opposed to FHSCs \citep{Maureira2020ALMACandidates}.

In this paper, we present VLA (Very Large Array) molecular line observations using the dense gas tracers NH$_3$(1,1) and NH$_3$(2,2), as well as 3mm NOEMA (Northern Extended Millimeter Array) dust continuum observations, of the FHSC candidate CB 17 MMS at a resolution of $\sim$4.5"(1120 au) and $\sim$2.5"(625 au), respectively. This candidate is located within the Bok Globule CB 17 at an estimated distance of 250$\pm{50}$ pc \citep{Launhardt2010LookingCores}. The Bok globule hosts the Class I protostar CB 17 IRS and was resolved into two 1.3 mm peaks (CB 17 SMM1 and SMM2) in single-dish observations with the IRAM 30 m telescope, separated by 5,000 au \citep{Launhardt2010LookingCores}. The CB 17 SMM1 peak is also well traced by Herschel SPIRE observations which show a rounded structure at this location.

The FHSC candidate CB 17 MMS was identified later by \citet{Chen2012SUBMILLIMETERCORE} based on the detection of a faint point source in SMA 1.3mm continuum observations with a resolution of $\sim$3”. The location of the 1.3 mm SMA peak is offset by $\sim$1,500 au from the 1.3 mm peak seen in the IRAM 30 m observations \citep{Schmalzl2014TheEPoS}. \cite{Chen2012SUBMILLIMETERCORE} show that the point source is undetected at wavelengths $\leq70$ $\mu$m, has a low bolometric luminosity ($\leq$ 0.04 L$_{\odot}$) and low bolometric temperature of about 10 K \citep{Chen2012SUBMILLIMETERCORE}, all in agreement with a very early evolutionary stage, such as a FHSC. CB 17 MMS was also found to drive a low velocity CO (2-1) molecular outflow ($\sim$2.5 km s\textsuperscript{-1} in SMA observations \citep{Chen2012SUBMILLIMETERCORE}), in agreement with MHD simulations of the FHSC stage (\citealt{Machida2008HighCloud}; \citealt{Tomida2010RADIATIONOUTFLOW}). However, \citet{Chen2012SUBMILLIMETERCORE} note that the outflow emission could come from the nearby Class I instead and that uncertainties in factors affecting outflow morphology and other physical properties necessitate further observations to determine the nature of this source definitively. This study reveals the lack of compact emission in new 3 mm sensitive NOEMA observations and provides evidence from molecular line observations that the structure is consistent with a marginally bound starless core. 


This paper is organized as follows: in Section \ref{sec:obs} we describe the observations and data reduction. Section \ref{sec:results} presents the general results prior to the  spectral fitting, including the dust continuum emission and the ammonia integrated intensity maps. In Section \ref{sec:analysis} we discuss the ammonia line fitting procedure and property maps produced by these fits and offer analysis of these properties with reference to the evolutionary state of the sources. Section \ref{sec:discuss} discusses the column density and angular momentum profiles and their interpretations with regards to the kinematics of the core. Finally, we present our conclusions in Section \ref{sec:conclu}.

\section{Observations} \label{sec:obs}

\begin{deluxetable*}{ccccc}
\tablenum{1}
\tablecolumns{5}
\tablecaption{\\ Map Parameters.\label{tab:sources}}
\tablehead{
\colhead{Map} &  \colhead{Rest Frequency\textsuperscript{a}} & \colhead{Synthesized Beam} & \colhead{P.A.} & \colhead{rms\textsuperscript{b}} \\
\colhead{} & \colhead{(GHz)} &  &  & \colhead{(mJy beam\textsuperscript{-1})}
}
\startdata
NH\textsubscript{3}(1,1) & 23.694495 & 5.0" x 4.3" & 54.22$^{\circ}$ & 6.2\\
NH\textsubscript{3}(2,2) & 23.722733 & 5.0" x 4.0" & 54.16$^{\circ}$ & 6.0\\
Continuum & 82 & 2.9" x 2.4" & 51.00$^{\circ}$ & 0.008\\
\hline
\enddata
\linespread{1.0}\selectfont{}
\tablecomments{
\textsuperscript{a}\cite{Ho1983InterstellarAmmonia}. \textsuperscript{b}The RMS for the molecular lines is measured using channels that are 0.05 km s\textsuperscript{-1} wide. \textsuperscript{c}
}
\end{deluxetable*}

\subsection{VLA observations}

VLA observations (Project ID 14A-362) were conducted with 27 antennas in the D configuration, 
which provides baselines from 0.035 to 1.03 km, resulting in a largest recoverable scale of $\sim$60" (15,000 au). Observations were conducted on July 8, 2014 with a total on-source integration time of 2.4 hours.
Each track included observations of a flux, bandpass, and science source-phase calibrator at the beginning, beginning or end, and intermittently throughout the track, respectively. For these observations, 
0542+498 was selected for flux calibration, J0319+4130 for bandpass calibration, and J0359+5057 for phase calibration.

Multi-line observations were obtained with the K-band (central frequency $\sim$22.3 GHz) 
receiver with 14 spectral windows. Eight of these windows have a width of 2 MHz, with dual polarization, 512 channels, and velocity coverage and resolution of 25 and 0.05 km s$^{-1}$, respectively. 
For the purpose of this paper, we focus on the NH$_3$(1,1) and (2,2) transitions (at 23.69 and 23.72 GHz respectively), which were
covered with two 4 MHz sub-bands each. Each window had dual polarization, 1024 channels, and velocity coverage and resolution of 50 and 0.05 km s$^{-1}$, respectively. 
NH\textsubscript{3} is a particularly good tracer of dense gas and allows observations of the central core region immediately surrounding the FHSC candidate. 
The software CASA (Common Astronomy Software Application; \citealt{McMullin2007CASAApplications}) 
was used for both calibration and imaging. Calibration of the raw visibility data was done using the standard VLA calibration pipeline.

We use \textit{tclean} in CASA 5.6.2 to image the calibrated visibilities. We employed the multiscale deconvolver  with scales of 0, 7, and 21 pixels, with a pixel cell size of 0.6\arcsec. These scales correspond to a point source, a beam, and three times the beam. Additional cleaning parameters used include a uv-taper of 2.5\arcsec and natural weighting. The final beam sizes and rms of the resulting maps are given in Table \ref{tab:sources}.

\subsection{NOEMA observations}

NOEMA 3mm single-pointing observations (project ID S20AD) were conducted using the C and D configurations between June 14th and October 19th 2020, with 10 antennas. The combined baseline range was 24 to 344 m, corresponding to a maximum recoverable scale\footnote{Calculated as $0.6\times\lambda$/B$_{min}$.} of $\sim$17" (4,250 au). The total time on source was 4.1 and 3.0 hours in C and D, respectively. The flux and bandpass calibrators were MWC349 and 3C454.3, respectively. The phase/amplitude calibrators were J0359+600 and 0355+508. 

Observations were carried with the wideband correlator PolyFix resulting in two 7.7 GHz-wide sidebands, with channel spacing of 2 MHz which were used for the continuum, after spectral line flagging. The upper sideband (usb) and lower sideband (lsb) were centered at $\sim$90 GHz and $\sim$74 GHz, respectively. 
Additionally, the setup included 27 high-resolution chunks with channel spacing of 62.5 kHz for lines, within the upper and lower sidebands. In this work, we focus on the continuum results only. 

Data calibration were performed with the GILDAS\footnote{\url{https://www.iram.fr/IRAMFR/GILDAS}} software CLIC using standard procedures with the observatory provided pipeline (\citealt{2005Pety}; \citealt{2013Gildas}). We use the GILDAS software MAPPING for imaging using the {\it multi} deconvolution algorithm, natural weighting and a user-defined mask. We use the task {\it uv\_merge} to combine the visibilities from the lsb and usb. In this task, a spectral index can be inputted to correct the flux of one of the uv-tables so that it is on the same scale as the flux of the other. We did not apply any correction, i.e., we use a spectral index of 0. The rest frequency  and half power primary beam of the combined continuum are 82.0 GHz and 61.5", respectively. The synthesized beam and rms of the final continuum map is 2.9"$\times$2.4"(P.A.$=$51$^{\circ}$) and 8.2 $\mu$Jy beam$^{-1}$, respectively. 




\section{Results} \label{sec:results}
\subsection{3mm Continuum}
\label{sec:3mmcont}
\begin{figure}
  \includegraphics[width=0.5\textwidth]{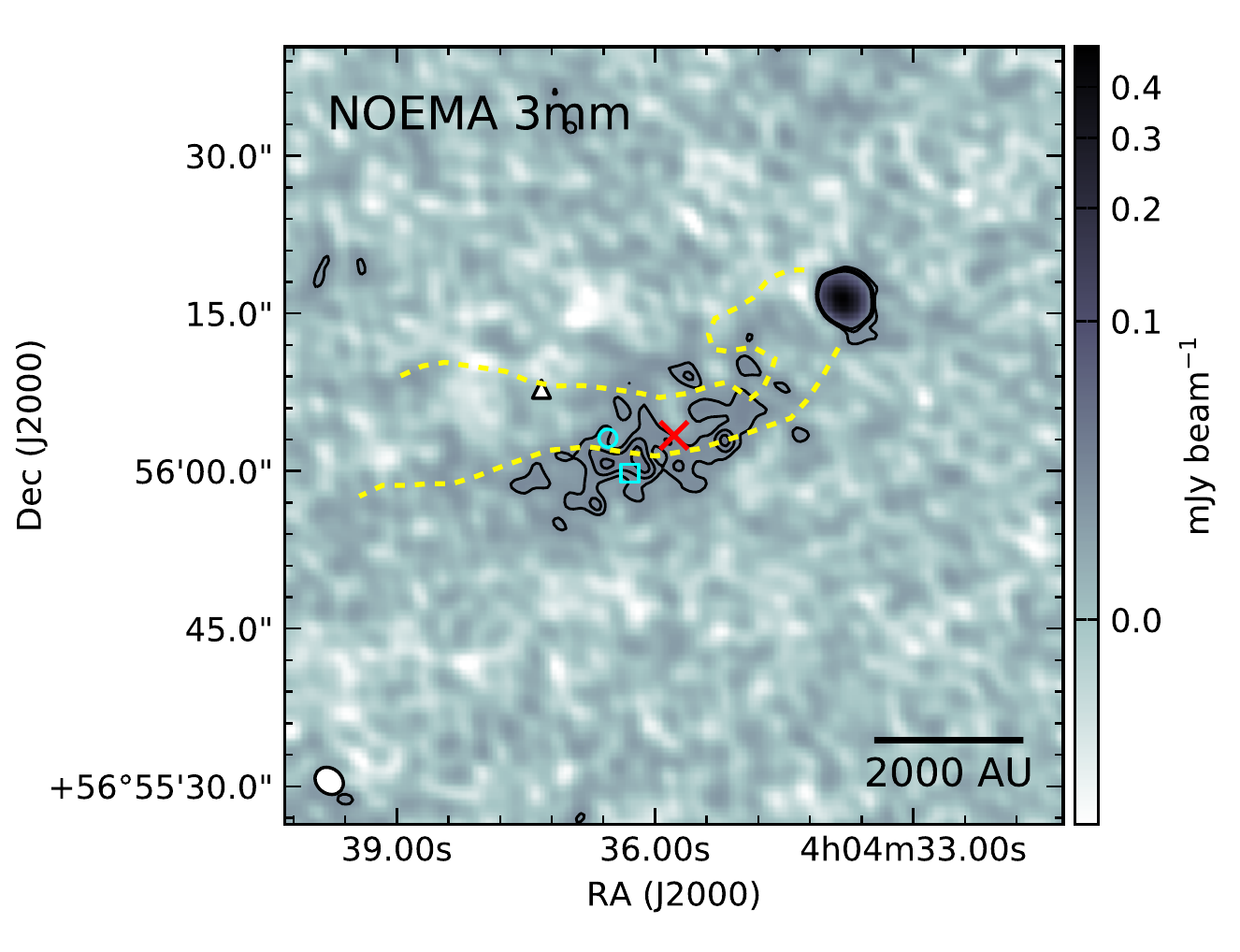}
2      \caption{NOEMA observations of the 3mm (82 GHz) continuum emission towards CB 17. The location of the 1.3mm compact source CB 17 MMS reported in SMA observations by \citealt{Chen2012SUBMILLIMETERCORE} is marked with a red cross. No compact source is detected at this location with a significance of at least $9\sigma$ (see text). The cyan circle marks the location of the 1.3 mm IRAM 30 m peak CB 17 SMM1 \citep{Launhardt2010LookingCores}. The cyan square marks the location of the NH$_3$ column density peak (Figure~\ref{fig:mom0_t_ncol}).
The compact emission to the northwest corresponds to the Class I source CB 17 IRS. Dashed yellow contours outline the CO(2-1) outflow emission extending to the southeast of CB 17 IRS (reproduced from \citealt{Chen2012SUBMILLIMETERCORE}). The white triangle marks the location of high velocity dispersion observed in NH$_3$  (Section~\ref{sec:dispersion}).
Black Contours are drawn at 3, 4 and 5$\sigma$. The beam is shown in the bottom left corner.  } \label{fig:3mm_cont}
\end{figure}

Figure~\ref{fig:3mm_cont} shows the combined usb and lsb 3mm NOEMA observations centered on the location of the 1.3 mm compact source detected in SMA observations and proposed as the FHSC candidate CB 17 MMS \citep{Chen2012SUBMILLIMETERCORE}. Our observations do not detect a compact source towards this location, marked with a red cross. The emission around the location of the candidate appears to be extended and thus, resolved out by our observations. There are small scale 1 to 2$\sigma$ fluctuations over the 3$\sigma$ extended emission, which can be expected in low-brightness extended emission. The reported peak flux of the 1.3 mm unresolved compact source in \cite{Chen2012SUBMILLIMETERCORE} is 3.3 mJy beam$^{-1}$ (6$\sigma$ detection), observed with a resolution comparable to ours. The new NOEMA observations were designed to be sensitive enough to obtain a clear continuum detection of the candidate based on the 1.3 mm SMA observations in \cite{Chen2012SUBMILLIMETERCORE}. Extrapolating the 1.3 mm flux using a spectral index of 3.7 (i.e., a ISM-$\beta$ dust emissivity index of 1.7) provides a strict lower limit to the expected peak flux at 3 mm of 0.07 mJy. Given the rms of the observations at 0.008 mJy/beam (Figure~\ref{fig:3mm_cont}), we expected a detection with a S/N of 9. A lower spectral index, expected when the emission is optically thick or in the presence of larger grains, would result in an even higher expected S/N. Similar results are obtained when analyzing the usb and lsb continuum emission separately. Based on these observations, we can rule-out the presence of a compact source at the location of the proposed FHSC candidate CB 17 MMS with a high significance. Thus, the new continuum observations indicate that CB 17 MMS is starless. \\

These results are similar to the case of a previous FHSC candidate (L1448 IRS2E), also proposed based on the detection of a 6$\sigma$ point source at 1.3mm in SMA observations \citep{Chen2010L1448Core}. The point source was later undetected at both 1.3mm SMA and 3mm ALMA observations which had sufficient sensitivity \citep{Stephens2018MassRelease,Maureira2020ALMACandidates}. Similar to CB 17, a nearby bright compact source was also present in the field of L1448 IRS 2E, which could have lead to artifacts being left over from the cleaning process in the case of both candidates. Such artifacts, combined with existing extended weak emission in the area and the neighbor outflow emission, may have led to their misclassification as FHSCs. Additionally, these FHSC candidates were also proposed based on observed CO(2-1) outflow emission related to the presumed compact source. In the case of CB 17 MMS, Figure~\ref{fig:3mm_cont} shows with dashed yellow contours the location and extent of the CO (2-1) emission to the southeast of CB 17 IRS reported in the SMA observations by \cite{Chen2012SUBMILLIMETERCORE}. The outflow emission changes direction near the location  of the presumed FHSC candidate and is seen just above the high NH$_3$ column density region (Figure~\ref{fig:mom0_t_ncol}). We believe that, similar to L1448 IRS 2E, the CO outflow emission thought to  originate in the FHSC candidate outflow, corresponds to outflow emission from the nearby protostar, deflected by the dense ambient material \citep{Maureira2020ALMACandidates}. \\

The only clear compact source detection in the field corresponds to the Class I source CB 17 IRS towards the northwest. We fit the compact source with a Gaussian (after applying primary beam correction) and we obtained an integrated and peak flux of 0.71$\pm$0.03 mJy and 0.65 $\pm$ 0.02 mJy beam$^{-1}$, respectively.


\subsection{NH$_3$ Integrated Intensity} \label{subsec:mom0}

\begin{figure*}
\gridline{\fig{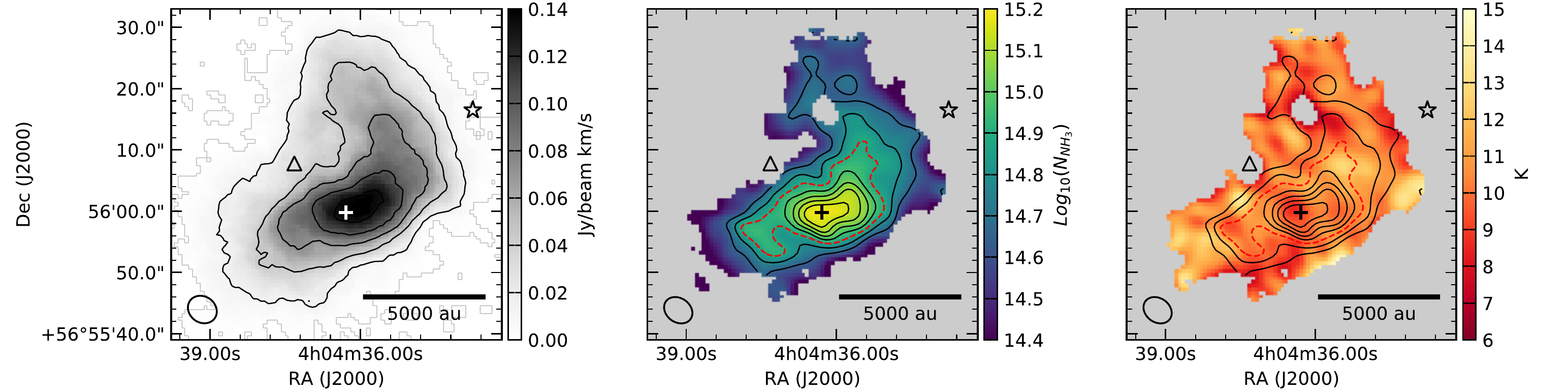}{1\textwidth}{}}
\caption{Integrated intensity map of NH$_3$(1,1) (left), NH$_3$ column density map (center), and NH$_3$ derived kinetic temperature map (right). The peak column density is marked by the + sign. The Class I source CB 17 IRS ($\bigwhitestar$) is marked in the upper right part of the panel. The region of high NH$_3$ velocity dispersion, which can be seen in Figure \ref{fig:vel_sigma}, is marked by the $\Delta$ symbol. The black oval in the bottom left corner indicates the synthesized beam. Left: Black contours are drawn at 10\%, 30\%, 50\%, 70\%, and 90\% of the peak integrated intensity (integrated intensity includes the satellite components). Center, right: Black contours are drawn at 30\%, 40\%, 60\%, 70\%, 80\%, and 90\% of the peak column density. The red dashed contour indicates the 50$\%$ peak column density.  
\label{fig:mom0_t_ncol}}
\end{figure*}

Given that we do not detect a compact source in the direction of CB 17 MMS with the 3mm NOEMA observations, we will henceforth refer to the dense starless region that we trace with NH\textsubscript{3} and discuss below as CB 17 SMM1. The SMM1 peak is closest to the NH\textsubscript{3} peak we observe with the VLA (see Figure~\ref{fig:3mm_cont}) and shows a similar rounded structure in the Herschel SPIRE observations. The left panel of Figure \ref{fig:mom0_t_ncol} shows the NH\textsubscript{3}(1,1) integrated intensity map of CB 17 SMM1, which includes the satellite components. Satellites refer to the NH\textsubscript{3}(1,1) hyperfine pattern, which includes 4 satellite lines and a main line. This structure is demonstrated by the spectrum in Figure \ref{fig:spec_model}. Only pixels above 3$\sigma$ were considered.
The emission exhibits a roundish shape measuring approximately 44" x 27" at the 10$\%$ peak emission contour, this corresponds to roughly 11,000 x 6700 au. The position angle of the major axis is 135$\pm$1.1$^{\circ}$ (measured East from North).
The position of the presumed 1.3mm compact source  \citep{Chen2012SUBMILLIMETERCORE}, lies at the edge of the contour demarcating the 90$\%$ value of the peak NH$_3$(1,1) integrated intensity. The nearby Class I source CB 17 IRS, located $\sim$21"(or 5,250 au) Northwest of CB 17 MMS and marked with a star in Figure~\ref{fig:mom0_t_ncol}, is not being traced by the ammonia. The integrated intensity map of NH\textsubscript{3}(2,2), showing a similar but weaker emission as the NH\textsubscript{3}(1,1), can be found in the Appendix (Figure \ref{fig:mom0_22}). 

\section{Molecular line analysis}\label{sec:analysis}

\subsection{Line Fitting} \label{subsec:fit}

The cold-ammonia model from the \texttt{python} package \texttt{pyspeckit} \citep{Ginsburg2011PySpecKit:Toolkit} was used to derive the kinematics and physical properties towards CB 17 from the NH\textsubscript{3}(1,1) and (2,2) emission. This package fits Gaussian profiles to the hyperfine components of the ammonia rotational transitions
using a model developed in \cite{Rosolowsky2008AnPerseus} and \cite{Friesen2009TheOphiuchus}, and relies on the theoretical framework established in \cite{Mangum2015HowDensity}. The free parameters of the fit are the fraction of NH\textsubscript{3} in the ortho state, the total NH\textsubscript{3} column density, the kinetic temperature (\textit{T}\textsubscript{K}), excitation temperature (\textit{T}\textsubscript{ex}), the line-of-sight velocity (\textit{v}\textsubscript{LSR}), 
and the velocity dispersion (\textit{$\sigma_v$}). In our fitting, for each pixel all parameters were left free except the fraction of NH\textsubscript{3} in the ortho state, which was fixed at 0.5, i.e., assuming the local thermodynamic equilibrium (LTE) ortho-para ratio value of 1. The model assumes the same  \textit{T}\textsubscript{ex} temperature for all hyperfines as well as for the (1,1) and (2,2) lines. The model derives \textit{T}\textsubscript{K} from the rotational temperature by analytically solving the equations of detailed balance described in \citet{Swift2005AL1551}. See Section 3.1 of \cite{Friesen2017Belt} for a more detailed derivation of the cold-ammonia model used in \texttt{pyspeckit}. 

\begin{figure*}
\gridline{\fig{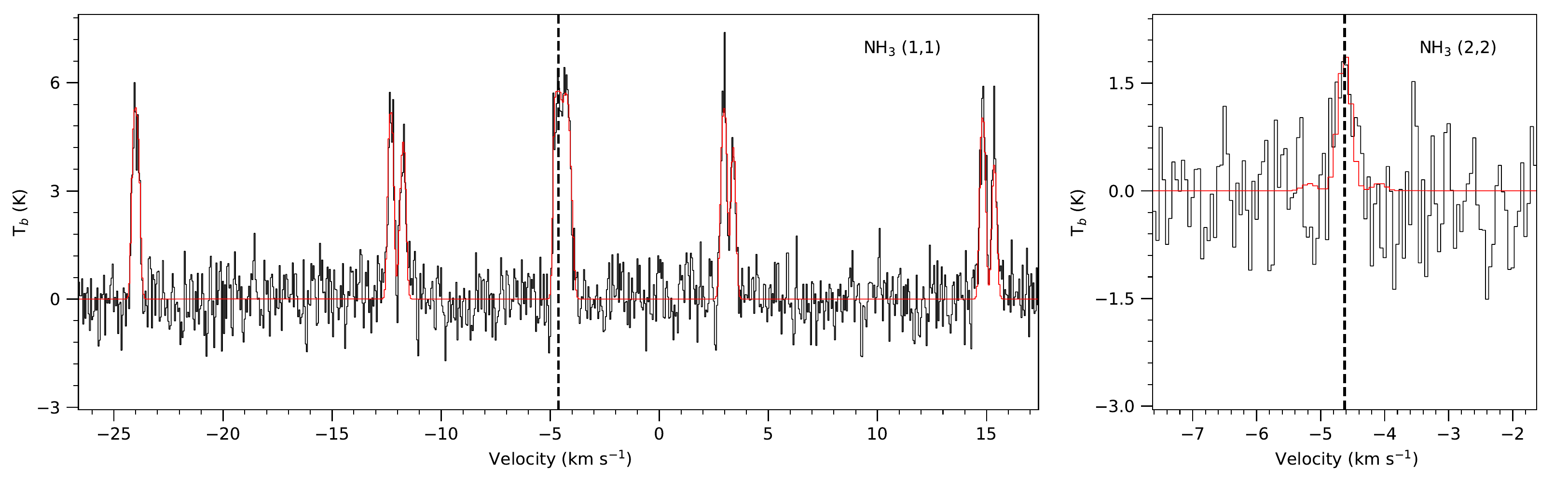}{1\textwidth}{}}
\caption{NH$_3$(1,1) (left) and NH$_3$(2,2) (right) spectrum at the derived column density peak. Observations are in black while the fit results are in red. The x-axis corresponds to LSR velocities. Vertical dashed lines mark the central velocity ($v_{LSR}=-4.63$ km$s^{-1}$) derived from the hyperfine fit at this position (Section~\ref{subsec:fit}).
\label{fig:spec_model}}
\end{figure*}

Fits were attempted only for those pixels for which the NH\textsubscript{3}(1,1) signal-to-noise ratio was above 5. After the fits were completed, we performed masking of any pixels for which the resultant error on an individual parameter was zero or the ratio between the parameter value and error was less than 3. This was done individually for each parameter. Further, we remove for \textit{T}\textsubscript{K} and \textit{T}\textsubscript{ex} all pixels for which the error was $\geqslant$ 2 K, and these same pixels were removed from the final column density map. The latter results in removal of sets of pixels at the edge of the integrated intensity map which had the largest errors in these parameters. The cutoff of 2 K was chosen to remove pixels where uncertainty was larger than the typical parameter variations seen in the map as it would be hard to detect any real variation in temperature if the errors in the pixels are larger than the variation itself. In the case of the velocity and velocity dispersion, the errors are typically much smaller than the variations throughout the maps, as these parameters are constrained more robustly in the fit in comparison with \textit{T}\textsubscript{K} and \textit{T}\textsubscript{ex}. Hence, for the velocity and velocity dispersion maps we did not apply any additional masking. This is why these are more extended than the temperature and column density maps (Fig. \ref{fig:vel_sigma}). An example of the spectra for NH\textsubscript{3}(1,1) and NH\textsubscript{3}(2,2) with the corresponding model is shown in Figure~\ref{fig:spec_model}. The spectra corresponds to those extracted at the position of the column density peak.

A summary of the values obtained from the resulting fits can be found in Table \ref{tab:map}. The resultant maps for NH\textsubscript{3} column density and kinetic temperature are shown in Figure \ref{fig:mom0_t_ncol} while the central velocity and velocity dispersion are shown in Figure \ref{fig:vel_sigma}. The excitation temperature map is shown in the right panel of Figure \ref{fig:vel_sigma}. 


\begin{deluxetable*}{cccc}
\tablenum{2}
\tablecolumns{4}
\tablecaption{\\Map Properties.\label{tab:map}}
\tablehead{
\colhead{Region} & \colhead{NH$_3$ Column Density} & \colhead{Temperature ($T_{kinetic}$)} &  \colhead{Velocity Dispersion ($\sigma_v$)}\\
\colhead{} & \colhead{(1 x 10\textsuperscript{14} cm\textsuperscript{-2})}  &
\colhead{(K)} & \colhead{(km s\textsuperscript{-1})}
}
\startdata
Total Core & 5.8 (2.8) & 10.4 (1.1) & 0.09 (0.02)\\
Inner Core & 10 (2.2) & 10.5 (0.8) & 0.1 (0.01)\\
NH\textsubscript{3} Column Density Peak & $15\pm{1.0}$ & $8.9\pm{0.4}$ & $0.1\pm{0.003}$\\
\enddata
\linespread{1.0}\selectfont{}
\tablecomments{Total Core and Inner Core values are given as Mean (Standard Deviation). Median values are slightly lower (within 1$\sigma$) than mean values. NH\textsubscript{3} Column Density Peak values are given as the value and error of the individual pixel. These properties were obtained by fitting the spectrum in each pixel (see Sec. \ref{subsec:fit}). We define the "Inner Core" as the region within 50\% peak column density (see Sec. \ref{subsec:coldensity}).
}
\end{deluxetable*}

\subsection{Column Density} \label{subsubsec:temp_ncol}

The column density increases towards the center, closely following the integrated intensity map (Figure~\ref{fig:mom0_t_ncol}). The mean NH\textsubscript{3} column density and NH\textsubscript{3} column density peak are 5.8 x 10$^{14}$ cm$^{-2}$ and 1.5 x 10$^{15}$ cm$^{-2}$, respectively (mean value is also reported in Table \ref{tab:map}). To explore the properties of the central area, we designate an `inner core' region within the 50$\%$ peak column density contour, which is demarcated via the red dashed contour on the map. In the following section we consider the mass and volume density of both the total core and the inner core. We discuss the column density profile in greater depth in section \ref{subsec:coldensity}.


\subsubsection{Mass and Density from ammonia}\label{subsubsec:mass}


We estimate the mass and volume density of the core using the column density from our earlier fit (Section~\ref{subsec:fit}) considering the entire mapped structure and the inner core region. For this, we use a fractional abundance with respect to H$_2$ of 10\textsuperscript{-8} for NH$_3$, consistent with the mean value reported by \cite{Friesen2017Belt} for NGC 1333 and B18 in Perseus and Taurus, respectively. However, there are a range of potential abundances which could be used. For example, \citet{Seo2015ANCORES} also analyze NH\textsubscript{3} masses of filaments L1495-B218 in Taurus using an abundance of $1.5\times10^{-9}$, which they verify by comparing the derived masses from the NH\textsubscript{3} column density and abundance to the masses estimated from the dust continuum and  the virial masses. We verify our abundance ratio using the central H$_2$ column density for CB 17 derived from lower resolution continuum observations ($\sim$3$\times10^{22}$ cm$^{-2}$, \citealt{Schmalzl2014TheEPoS}) and our average NH$_3$ column density, which leads to an abundance of $1.7\times10^{-8}$ (in agreement within a factor of two of the assumed abundance). The spatial resolution of the \citet{Schmalzl2014TheEPoS} observations is of a similar scale to the NH\textsubscript{3} emission region we detect (25" or 6250 au beam). 
Considering all of the pixels in the column density map and a mean molecular weight per hydrogen molecule of $\mu_{H_2}$=2.74   \citep{Tanner2011TheMm}, we obtain a 
total mass of  1.70 ($\times10^{-8}/X[$NH$_3]$) $M_\odot$. This method is similar to that used in \citet{Seo2015ANCORES}. Our estimation agrees with the $2.3\pm{0.3} M_\odot$ value estimated in \cite{Schmalzl2014TheEPoS} derived from dust Herschel and single-dish observations that traced a larger  structure (by a factor of 4) with a radius of 18 x 10\textsuperscript{3}au. It is also consistent with the masses of NH\textsubscript{3} leafs (structures containing individual intensity peaks) in Taurus filaments, which range from 0.05 to 4.3 $M_\odot$ \citep{Seo2015ANCORES}.

\begin{deluxetable*}{ccccccc}
\tablenum{3}
\tablecolumns{8}
\tablewidth{0.9\textwidth}
\tablecaption{\\ Properties Derived from Ammonia.\label{tab:mass}}
\tablehead{
\colhead{Total Mass} & \colhead{Inner Mass} & \colhead{Total Volume Density} & \colhead{Inner Region Size} & \colhead{Inner Volume Density} & \colhead{Magnitude} & \colhead{Direction}\\
\colhead{$(M_\odot)$} & \colhead{$(M_\odot)$} & \colhead{(10\textsuperscript{6} cm\textsuperscript{-3})} & \colhead{(au)} & \colhead{(1 x 10\textsuperscript{6} cm\textsuperscript{-3})} & \colhead{(km s\textsuperscript{-1} pc\textsuperscript{-1})} & \colhead{deg.~ E of N}
}
\startdata
$1.70$ & $0.63$ & 1.0 & 2,950 x 1,180 & 1.9 & $5.56\pm{0.07}$ & $-58.3\pm{0.7}$\\
\enddata
\linespread{1.0}\selectfont{}
\tablecomments{
We assumed a NH\textsubscript{3} fractional abundance with respect to H$_2$ of $10^{-8}$ for all estimates \citep{Friesen2017Belt}. The Total Mass and Total Volume Density estimates are calculated for the entire column density map. The Inner Mass and Inner Volume Density estimates are calculated within 50\% of the column density peak (demarcated by the red contour in Figure \ref{fig:mom0_t_ncol}). The region size corresponds to the semi-major and semi-minor axes of this Inner Region.}
\end{deluxetable*}

To obtain a lower limit estimate of the volume density, we consider the volume of an ellipsoid and assumed that the third axis (along the line of sight) is equal to the major axis of the observed emission. 
The semi-major and semi-minor axes of the main structure correspond to $\sim$4,800 and $\sim$2,300 au, respectively (the axis lengths are visually determined). This results in a gas volume density estimate of $\sim$1$\times10^6$ cm\textsuperscript{-3} ($\times10^{-8}/X[$NH$_3]$). For the 50$\%$ of the peak column density region, with semi-major and semi-minor axes of 2,950 and 1,180 au, respectively (axis lengths are again visually estimated), we derive a mass of 0.63 $M_\odot$ ($\times10^{-8}/X[$NH$_3]$) and a higher corresponding gas volume density of $\sim$2$\times10^6$ cm\textsuperscript{-3} ($\times10^{-8}/X[$NH$_3]$). Even considering uncertainties in the abundance of a factor of a few, the estimated central density is close to or above 10$^6$ cm$^{-3}$. We later give another estimate of volume density based on the fit of the column density profile which is consistent with these estimates within a factor of a few (see Table \ref{tab:fit_params} and Section \ref{subsec:coldensity}). This value is consistent with the detection of N$_2$D$^+$(3-2) in SMA observations \citep{Chen2012SUBMILLIMETERCORE}, which has a  critical density of 3$\times$10$^6$ cm$^{-3}$. In addition, the integrated intensity peak of the N$_2$D$^+$(3-2) closely matches the location of the NH$_3$ column density peak and is unrelated to any peak in the NH$_3$ (2,2) emission (Figure~\ref{fig:mom0_22}). This suggests that the NH$_3$ column density peak is indeed tracing the higher density region within the core and not a region with enhanced temperature. In comparison, \citet{Schmalzl2014TheEPoS} and \citet{Lippok2013Gas-phaseCores} report central volume densities of only $1.3-2.3 \times 10^5$ cm\textsuperscript{-3}, using several dust emission maps convolved to a common resolution that is a factor of 5  lower than the NH$_3$ maps presented here. The results discussed above are summarized in Table~\ref{tab:mass}. 



\subsection{Temperature}

Figure \ref{fig:mom0_t_ncol} includes a map of  the kinetic temperature obtained through the fit of  spectra of  NH$_3$ (1,1) and (2,2) towards each pixel,  described in Sec.\ref{subsec:fit}. The map shows the average temperature along the line-of-sight for each pixel with significant NH$_3$ (1,1) and (2,2) emission in the CB 17 SMM1 core. We calculate the median temperature (and standard deviation) to  be  10.4 $\pm$1.1 K (see also Table \ref{tab:map}), derived using a KDE density distribution from all pixels in the temperature map (Figure~\ref{fig:mom0_t_ncol}). This value matches the dust temperature of 10.6 $\pm$0.3 K found in a inner isothermal region within 4,000 au \citep{Schmalzl2014TheEPoS}, derived from Herschel and single-dish continuum observations and modeled considering gradients in volume density and temperature, i.e., not from line-of-sight averages. Line-of-sight average results in a larger central dust temperature of 14.2$\pm$0.7 K \citep{Schmalzl2014TheEPoS}.



\subsection{Central Velocity}\label{subsubsec:vel}
\begin{figure*}
\gridline{\fig{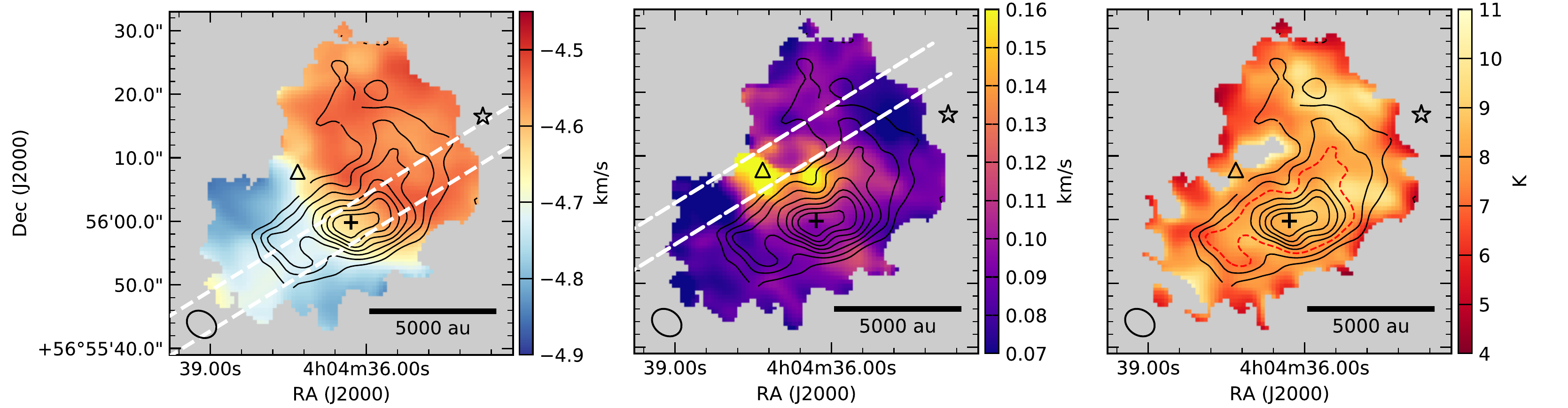}{1\textwidth}{}}
\caption{Velocity (left), line-width (center), and excitation temperature (right) maps of CB 17 MMS. The peak column density is marked by the + sign. The Class I source CB 17 IRS ($\bigwhitestar$) is marked in the upper right part of the panel. The region of high velocity dispersion is marked by the $\Delta$ symbol. The black oval in the bottom left corner indicates the synthesized beam. Black contours are drawn between 30\% and 90\% of the peak column density in increments of 10\%. The velocity and line-width maps show the values obtained by fitting a Gaussian profile to the molecular line spectra as discussed in \ref{subsec:fit}. The cuts used to make the position-velocity diagrams in Figure \ref{fig:pv} are marked by the white dashed lines. The cut in the velocity map, centered on the peak NH\textsubscript{3} column density, corresponds to the left panel of Figure \ref{fig:pv} while the cut in the line-width map, centered on the high velocity dispersion region, corresponds to the right panel of Figure \ref{fig:pv}.
\label{fig:vel_sigma}}
\end{figure*}

The velocity map for the core can be found in Figure \ref{fig:vel_sigma}. 
The core displays a clear velocity gradient increasing from Southeast to Northwest along the major axis of the structure. To measure the magnitude and direction of the systematic velocity gradient, we employ the methodology described in Section 2.1 of \citet{Goodman1993DenseGradients}. We begin by fitting the function for a linear velocity gradient in a plane,
\begin{equation} \label{eq:vel}
v_{LSR} = v_{0} + a\Delta\alpha + b\Delta\beta .
\end{equation} 
$\Delta\alpha$ and $\Delta\beta$ represent the radian offset in right ascension and declination while $a$ and $b$ are the gradient per radian projections on these axes. $v_0$ is the core systemic velocity. Given Equation \ref{eq:vel}, the magnitude of the gradient can be shown as 
\begin{equation} \label{eq:mag}
\mathcal{G} \equiv \mid\nabla v_{LSR}\mid = (a^2 + b^2)^{1/2}/D ,
\end{equation} 
where \textit{D} is the distance to the cloud. The gradient direction (measured East of North, in the direction of increasing velocity) is then given by
\begin{equation} \label{eq:angle}
\theta_\mathcal{G} = tan^{-1}\frac{a}{b} .
\end{equation} 
Based on this fit, we find the core systemic velocity $v_0$ is $-4.64\pm{0.01}$ km s\textsuperscript{-1}. The magnitude of the velocity gradient and its direction are $5.56\pm{0.07}$ km s\textsuperscript{-1} pc\textsuperscript{-1} and $-58.3\pm{0.7}^{\circ}$ east from north, respectively (Table \ref{tab:mass}). 

\begin{figure*}
\gridline{\fig{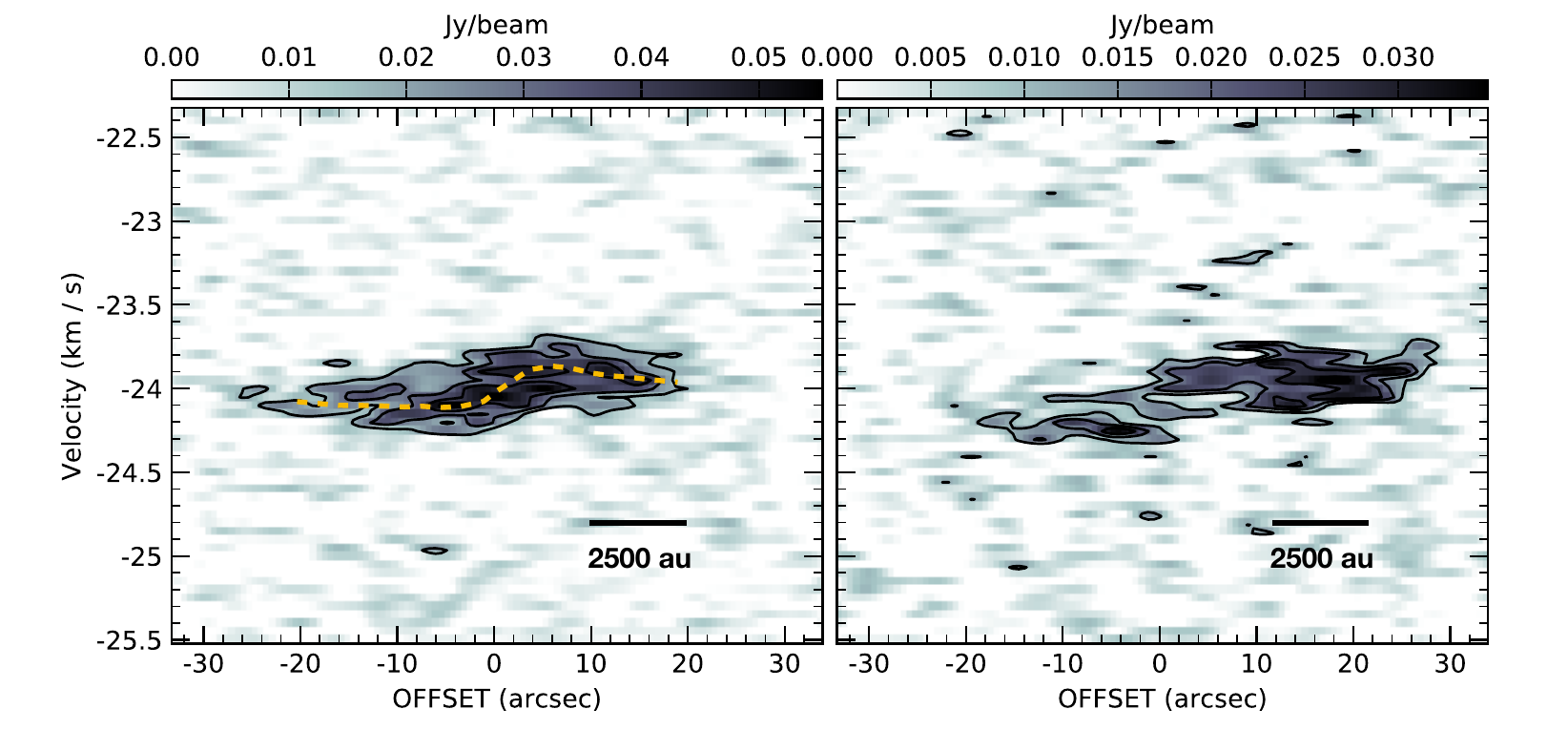}{0.9\textwidth}{}}
\caption{Position-velocity maps of the ammonia emission in CB17 SMM1 along the P.A of the derived velocity gradient, using only the satellite hyperfine component close to -24 km~s$^{-1}$ (see Fig.~\ref{fig:spec_model}).
Left: p-v diagram centered at the NH$_3$ column density peak. The dashed line highlights the sinusoidal or arc-like feature discussed in the text. Right: p-v diagram centered on the high velocity dispersion region (triangle marker in Figure \ref{fig:vel_sigma}). Black contours are drawn at 30\%, 50\%, 70\%, and 90\% of the peak emission. The direction and width of  the position-velocity diagrams are indicated by the white dashed lines in Figure \ref{fig:vel_sigma}. 
\label{fig:pv}}
\end{figure*}

\citet{Schmalzl2014TheEPoS} report a comparable gradient of $\sim$4.3$\pm{0.2}$ km s\textsuperscript{-1} pc\textsuperscript{-1} using the N\textsubscript{2}H\textsuperscript{+}(1-0) line. They find an arc-like or sinusoidal pattern in the position-velocity diagram of N\textsubscript{2}H\textsuperscript{+}(1-0) (see their Figure 7).
We find a similar  arc-like pattern in the position-velocity diagram for NH\textsubscript{3}(1,1) centered on the column density peak (see left panel in Figure \ref{fig:pv}).

Generally speaking, velocity gradients at core ($\sim$0.1 pc) and envelope scales (few $1,000$ au) are usually interpreted as the result of infall or rotational motions, but they could also arise from large-scale turbulence or convergent flow compression \citep{Burkert2000TurbulentProperties,Ballesteros-Paredes2006MolecularFormation,Pineda2019TheFormation}. In the case of CB 17 SMM1, the arc-like (or sinusoidal, \citealt{Schmalzl2014TheEPoS}) structure of the p-v diagram can be due to the interaction of the dense material with the CO outflow emission originating from the nearby CB 17 IRS Class I source. Interestingly, a similar arc-like shape was recently reported by \cite{Maureira2020ALMACandidates} towards the extremely young Class 0 protostar Cha-MMS1 using N$_2$H$^{+}$(1-0) observations (see their Figure 10 bottom right panel). The envelope of this protostar is also found to be interacting with a nearby outflow. In both cases the velocity gradient of the structure was closely aligned with the nearby outflow axis. Moreover, two other starless cores (B1-NE and B1-SW in Perseus), observed also in N$_2$H$^{+}$(1-0),  show similar arc-like structures in their p-v diagrams (\citealt{Chen2019InvestigatingGBT-Argus}, see their Figure 3). Similar to CB 17 SMM1 and Cha-MMS1, the velocity maximum in the p-v diagram for these sources was not the location of the higher-density peak, as traced by the integrated molecular emission. \citet{Chen2019InvestigatingGBT-Argus} suggest that these velocity structures might be due to convergence of two oblique gas flows. Similar convergent flow compression could arise in the case of CB 17 and Cha-MMS1, from the interaction between the nearby outflow 
and the dense core gas traced by the dust continuum and ammonia emission  (Figures~\ref{fig:3mm_cont} and~\ref{fig:mom0_t_ncol}).


\subsection{Velocity Dispersion}
\label{sec:dispersion}

Figure \ref{fig:vel_sigma} shows a map of the NH$_3$ velocity dispersion. Except for a small region on the eastern side of the  core, the velocity dispersion remains relatively constant. The median over the whole region is $\sigma_{v,med}=0.09$ km s$^{-1}$. This corresponds to a total line width (fwhm) of $\Delta$v $\sim$0.21 km s$^{-1}$, which is in agreement with the value obtained from only NH$_3$ thermal motions at a temperature of 10.4 K (our derived mean temperature, Sec.~\ref{subsubsec:temp_ncol}). 
These values and distribution are in agreement with the results   presented in \citet{Schmalzl2014TheEPoS} using the N$_2$H$^+$(1-0) line. 

Given that the derived temperatures are mostly uniform over the mapped region (Figure~\ref{fig:mom0_t_ncol}), the high-velocity dispersion region near the eastern edge of the core corresponds to an enhancement of non-thermal motions. The typical non-thermal NH$_3$ line width of $\Delta$v=0.33 km s$^{-1}$ \footnote{Corresponding to $\sigma=0.17$ km s$^{-1}$ in the map.} is calculated using our derived mean temperature value of 10.4 K (Section~\ref{subsubsec:temp_ncol}). This value and the H$_2$ thermal line width at this same temperature ($\Delta$v=0.45 km s$^{-1}$) leads to a H$_2$ non-thermal to thermal ratio of $\sim$0.7. Thus, the entire NH$_3$ mapped region, including the region with broad line widths, show sub-sonic non-thermal motions. These values are entirely in agreement with \cite{Schmalzl2014TheEPoS} observations of N$_2$H$^{+}$(1-0) which combined both single-dish and interferometric observations (IRAM 30m and PdBI). The good agreement also suggests that our line widths are not greatly affected by missing flux from large-scale structures  due to lack of single-dish observations. \\


The velocity dispersion map in Figure \ref{fig:vel_sigma} exhibits a region of high velocity dispersion on the eastern side of the core, marked by the $\Delta$ symbol in the figure. To investigate this region of broad line widths and its possible origin, 
we inspected position-velocity diagrams along the velocity gradient of the core and passing through the region (Figure \ref{fig:pv}). We find that the velocity gradient around this region is not due to the central velocity of the line shifting smoothly with position from blue-shifted to red-shifted. Instead there are two, fairly constant velocity components, which  overlap at the location at which the large line widths are observed. As the spectra were fitted with a single component, the local region in which the two components overlap naturally results in broad line widths. This type of velocity structure is in agreement with our previous interpretation of converging flows interaction near the center of the core. Both the region of broad line widths and the 1.3 mm IRAM peak are located within the CO(2-1) outflow emission from CB 17 IRS (Figure~\ref{fig:3mm_cont}). The region with broad line widths also corresponds to a region with weak emission and thus low column densities of NH$_3$. Therefore, the broad line widths are likely tracing the regions where the outflow is interacting with the dense core gas. Given that we do not see a high temperature associated with the broad line-width zone, any outflow impact would be tangential rather than direct.

\section{Discussion} \label{sec:discuss}
\subsection{Core structure and stability} \label{subsec:coldensity}
\begin{figure}
\gridline{\fig{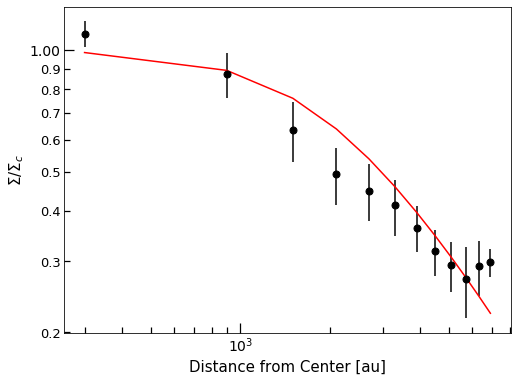}{0.45\textwidth}{}}
\caption{Azimuthally averaged column density distribution, derived from the NH$_3$ column density, assuming a NH$_3$ fractional abundance of 10$^{-8}$ with respect to H$_2$ and centered on the NH$_3$ column density peak. The solid red line shows the curve of the best-fit model (Table \ref{tab:fit_params}).
\label{fig:density_prof}}
\end{figure}

To better understand the structure and dynamical state of the core we performed a fit of the column density profile, presented in Figure \ref{fig:density_prof}. We use the same abundance ratio for NH\textsubscript{3} ($10^{-8}$) and mean molecular weight per hydrogen molecule ($\mu_{H_2}$=2.74) as we did for our previous mass and volume density estimates. The column density structure, particularly at inner regions of the core, can be an indicator of the evolutionary state of the core \citep{Ward-Thompson1994ACores,Bergin2007ColdFormation,Koumpia2020MappingTracers,Ward-Thompson2006AnFunction}. In prestellar cores, the characteristic “plateau” of slowly decreasing density, occurs within radii smaller than 2500-5000 au (\citealt{Andre1996Probing1689B.}; \citealt{Ward-Thompson1999TheCores};  \citealt{Bergin2007ColdFormation}). This feature is typically coupled with a power-law density decrease ($\sim$ $r^{-2}$) at larger radii (\citealt{Chandrasekhar1949TheFunction.}; \citealt{DiFrancesco2006AnProperties}). There are a number of different models used to fit such cores, including the original dual power law of \citet{Ward-Thompson1994ACores}, modified power laws of \citet{Tafalla2002SystematicCores}, and the truncated isothermal Bonnor-Ebert sphere model (\citealt{Bonnar1956BoylesInstability}; \citealt{Ebert1955UberTextabbildungen}). For our study, we employ the analytic model of \citet{Dapp2009AnCores}, which replicates the characteristics of both the Bonnor-Ebert sphere isothermal equilibria and non-equilibrium models, such as that of \citet{Larson1969NumericalProto-Star}. 

The \citet{Dapp2009AnCores} profile is defined as 
\begin{equation} \label{eq:profile}
\varrho(r) = \left\{
  \begin{array}{lr}
    \varrho_ca^2/(r^2+a^2) & : r\leq R\\
    0 & : r>R.
  \end{array}
\right.
\end{equation} 
This model is characterized by a central volume density $\varrho_c$, an outer radius $R$ at which point the core is truncated, and a parameter $a$ which fits the inner radius or plateau region. Integrating the volume density
along the line of sight and combining this expression with equation (\ref{eq:profile}) produces a closed-form expression of the column density 
\begin{equation} \label{col_density}
\Sigma(x) = \frac{2a^2\varrho_c}{\sqrt{x^2+a^2}} \text{arctan} \Bigg(\frac{\sqrt{R^2-x^2}}{\sqrt{x^2+a^2}}\Bigg) .
\end{equation}

Rewriting this in terms of the central column density $\Sigma_c$ with $\Sigma(x=0) \equiv \Sigma_c = 2a\varrho_c\text{arctan}(c)$ leads to
\begin{align} \label{BE_fit}
\Sigma(x) = & \frac{\Sigma_c}{\sqrt{1+(x/a)^2}} \nonumber \\
      & {} \times \Bigg[\text{arctan} \Bigg(\sqrt{\frac{c^2-(x/a)^2}{1+(x/a)^2}}\Bigg)\Bigg/\text{arctan}(c)\Bigg] ,
\end{align}
where $c \equiv R/a$. See \citet{Dapp2009AnCores} for a more detailed discussion of this derivation. 

\begin{deluxetable}{cl}
\tablenum{4}
\tablecolumns{2}
\tablewidth{0.95\textwidth}
\tablecaption{\label{tab:fit_params}}
\tablehead{\multicolumn{2}{c}{Column Density Profile Fit Parameters
}}
\startdata
$\Sigma_c$ & $(1.3 \pm 0.1) \times 10^{23}$ cm\textsuperscript{-2}\\
$\varrho_c$ & $(2.7 \pm 0.5) \times 10^{-18}$ g cm\textsuperscript{-3} \\
 & $(0.6 \pm 0.1) \times 10^{6}$ cm\textsuperscript{-3} \\
$r_0$ & 1800 $\pm$ 300 au\\
\enddata
\tablecomments{The outer radius r\textsubscript{out} was not constrained and was set to 30,000 au in this fit \citep{Schmalzl2014TheEPoS}. We assume an abundance of 10$^{-8}$ for NH$_3$ with respect to H$_2$ and a mean molecular weight $\mu_{H_2}$ of 2.74. 
}
\end{deluxetable}


We use this model to fit the azimuthally averaged H\textsubscript{2} column density profile of the core (Figure \ref{fig:density_prof}). 
The best-fit parameters are listed in Table \ref{tab:fit_params}. We find a central column density $\varrho_c$ of $0.6\pm0.1\times10^6$ cm$^{-3}$. Based on this fit we find an inner radius $r_0$ $\sim$1,800 au for the central flat region of the core, in agreement with the geometric mean of the radius containing 50\% of the peak column density. This inner radius size is also consistent with the radius range of the characteristic central flattening seen in prestellar cores (\citealt{Andre1996Probing1689B.}; \citealt{Ward-Thompson1999TheCores};  \citealt{Bergin2007ColdFormation}) and with the flat inner region derived using Equation 2 in \citet{Keto2010DynamicsCores}. 
The outer radius of the core is not well constrained by the fit. The limited extent of the ammonia emission does not provide enough data at large radii for the fit to determine the steepness of the density decrease in the outer regions of the core;   single-dish data (e.g., GBT) and mosaic observations would be needed for that. Nevertheless, the value of the outer radius does not affect our results of the inner radius or the central density. Hence, we fix this value at $3 \times 10^4$ au, which is identified as the outer radius from the fit performed by \citet{Schmalzl2014TheEPoS} based on Herschel and submm/mm continuum observations. 


\subsubsection{Stability Assessment}
\label{sec:density_profile}  
We use the results from the column density fit to evaluate the stability of the core, following \cite{Dapp2009AnCores}.  The inner radius \textit{a} is given by
\begin{equation} \label{inner_r}
a = k\frac{c_s}{\sqrt{G \varrho_c}} ,
\end{equation}
where \textit{k} is a proportionality constant, \textit{G} is the gravitational constant, and the isothermal sound speed $c_s$ is defined as $c_s = \sqrt{k_B T/\mu}$. We take the central volume density $\varrho_c$ from the fit of the central column density (where $\Sigma_c=2a\varrho_c\text{arctan}(c)$). \citet{Dapp2009AnCores} demonstrate that cores considered to be in equilibrium have a $k \approx 0.4$ while cores for which \textit{k} is closer to 1 are at risk of collapsing. Based on our fit, using a value of 10.4 K for the temperature, we find a \textit{k} value of 0.57. This indicates that the core is straddling the boundary between equilibrium and collapse.

We also investigate the stability of CB 17 SMM1 by using the virial parameter. We obtain the virial mass using the following equation:
\begin{equation} \label{m_vir}
M_{vir} = 5\sigma^{2}_{H_2}R/G ,
\end{equation}
which assumes a uniform density profile \citep{Bertoldi1992Pressure-confinedClouds}. The total velocity dispersion (thermal and non-thermal) $\sigma_{H_2}$ is estimated using the following:
\begin{equation} \label{sigma_H2}
\sigma^{2}_{H_2} = k_bT/\mu + (\sigma^{2}_{obs} - k_bT/m_{mol}),
\end{equation}
where $\sigma_{obs}$  corresponds to the observed velocity dispersion of the molecular line and $m_{mol}$ is the mass of the molecule \citep{Chen2007Stars}. Using the average observed velocity dispersion for $\sigma_{obs}$ we find a virial mass of $1.24 M_{\odot}$. The virial parameter $\alpha = M_{vir}/M$ is then 0.73. For a gravitationally bound system $\alpha \sim 1$ while in an unbound system $\alpha \gg 1$. Based on our $\alpha$ value of 0.73 and earlier derived k value we conclude that, at the scales at which we are observing the system, the core appears marginally bound. This is also consistent with the results from \citet{Schmalzl2014TheEPoS} that compare the gravitational energy with the rotational, turbulent and thermal energies. They conclude that the core becomes marginally bound at a radius from 8,000 to 18,000 au. 


\subsection{Central temperature}

\begin{figure}
\gridline{\fig{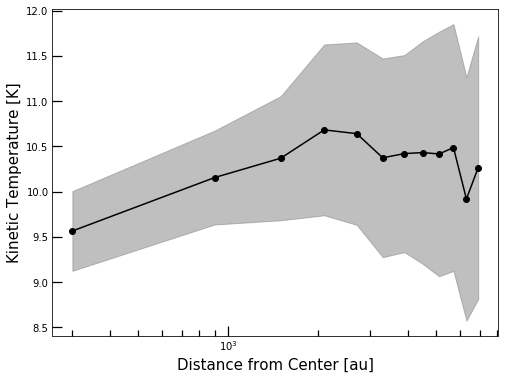}{0.45\textwidth}{}}
\caption{Azimuthally averaged kinetic temperature distribution centered in the NH$_3$ column density peak. Each annulus is approximately one beam size (1200au) in width. The errors, demarcated by the grey region, are determined by propagating the error in the fit for each individual pixel and averaging over each bin.
\label{fig:radial_temp}}
\end{figure}

Dense (starless) cores with a central density above $10^{5}$ cm$^{-3}$ are expected to exhibit a drop in their dust and gas temperature towards the center \citep{2005KetoDark}. 
Figure \ref{fig:radial_temp} shows the azimuthally averaged NH$_3$ temperatures as a function of distance from the NH$_3$ column density peak, where each annulus is approximately one beam size (1200 au) in width. The errors are determined by propagating the error from the fit  over each bin. The plot hints at a drop in temperature towards the center of about 1 K starting at about 2,000 au and reaching a central temperature of $9.5$ K. 

Central drops in line-of-sight averaged temperatures derived from NH$_3$ have been observed towards the evolved prestellar core L1544 from 6,000 down to 1,000 au scales \citep{Crapsi2007Observing1544} and the southern prestellar core in Ophiuchus D from 3,000 to 500 au scales \citep{Ruoskanen2011MappingAmmonia}. Both of these prestellar cores exhibit central temperatures lower than what we obtain for CB 17 ($\sim$6 and $\sim$7.5 K for L1544 and sourthern Oph D, respectively), close to the expected dust temperature in cores with a central density of $\sim10^{6}$ cm$^{-3}$ \citep{2005KetoDark}. On the other hand, the northern prestellar core in Ophiuchus D, shows a fairly constant temperature of 9 K, in better agreement with the values derived here for CB 17 SMM1 \citep{Ruoskanen2011MappingAmmonia}. As discussed in \citep{2008KetoDifferent}, at densities below $\sim$10$^{5}$ cm$^{-3}$ the gas cools down primarily through molecular line radiation keeping the gas temperature nearly isothermal, while above this threshold it starts to cool down mainly through collision with dust, resulting in a drop in the gas temperature towards the center. The two cores which shoe a drop temperature toward their central region have higher central densities close to or above 10$^6$ cm$^{-3}$, while Ophiuchus D North has a central density of the order of a few 10$^5$. The hint of a temperature drop seen in our derived temperature profile towards CB 17 supports our results from Sec.~\ref{subsec:coldensity}
that a volume density $\sim6\times$10$^{5}$ cm$^{-3}$ is present towards the very center of this core. 

For comparison, \cite{Keto2010DynamicsCores} provide theoretical density and gas temperature profiles for a starless core with a central density of 10$^{6}$ cm$^{-3}$. Their gas temperature profile drops from 11 K at $\sim$3,000 au down to 
8 K in the center, which is consistent with our results considering the uncertainties and that our measurement is a line-of-sight average. i.e., does not consider a temperature gradient towards the center which can result in an overestimation of the central temperature \citep{Schmalzl2014TheEPoS}. On the other hand, we can also consider the scenario that the gas temperature at the center of CB 17 remains around 9 K, even though the dust temperature is as low as 6-7 K, as expected in cores with up to 10$^{6}$ cm$^{-3}$ \citep{Keto2010DynamicsCores}. The difference between dust and gas temperature depends on dust properties such as dust grain size $a_{\rm eff}$ and cosmic ray ionization rate $\zeta$. Assuming spherical grains and that gas heating is due only to cosmic rays while cooling is only due to collision with dust grains, we can provide estimates of the typical dust grains sizes required to keep the gas temperature at 9 K, while having a dust temperature close to 6 K \citep{2019IvlevGasandDust}. Given a power-law distribution for sizes of dist grains, coagulation of grains reduces the total surface area of dust and hence, the coupling between gas and dust. This results in larger differences between dust and gas temperatures \citep{2019IvlevGasandDust}. Using equation 18 in \cite{2019IvlevGasandDust} for a temperature difference of 3 K and assuming a standard value of $\zeta = 10^{-17}$ s$^{-1}$, we derive a $a_{\rm eff} = 6$ $\mu$m. The expression in \cite{2019IvlevGasandDust} does not consider the growth of mantles on grains and the fact that they are likely not perfectly spherical, both of which should increase the coupling. Because of this, grains of a few tens $\mu$m in size are likely to be necessary for the scenario in which the central gas temperature remains around 9 K. Another possible scenario to explain the only marginal drop even though the core has a central volume density close to 10$^6$ is that the dust temperature is higher towards this source due to additional heating from the nearby protostar. Future observations at a higher resolution, or radiative transfer modeling can help to further constraint the density and temperature profile in this core.

\subsection{A starless core angular momentum profile} \label{sec:angmoment}

\begin{figure}
\gridline{\fig{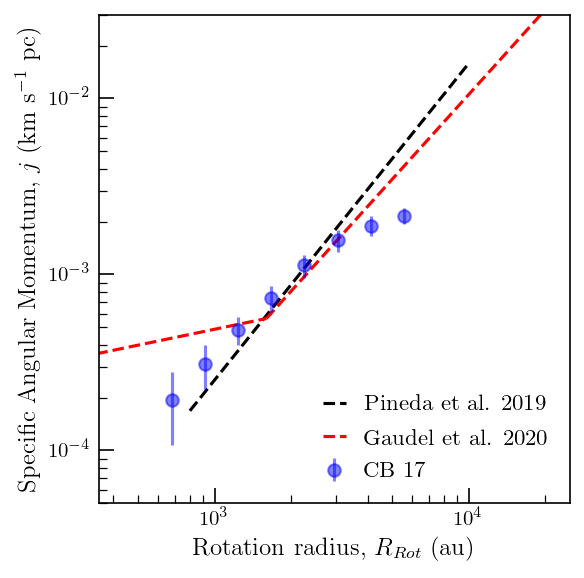}{0.45\textwidth}{}}
\caption{Radial profile of the specific angular momentum ($j = R_{rot}V_{rot}$) of CB 17. The dashed black line shows the best-fit power-law relation from \citet{Pineda2019TheFormation}. The dashed line shows the broken power-law model of the specific angular momentum from \citet{Gaudel2020AngularEnvelopes}. 
\label{fig:ang_moment}}
\end{figure}

The specific angular momentum radial profile may  be indicative of the evolutionary state of a core \citep{Pineda2019TheFormation}. Assuming symmetry around an axis, the specific angular momentum at a distance \textit{r} is described as $j(r) = R_{rot}V_{rot}$ where $R_{rot}$ is the rotation radius or impact parameter and $V_{rot}$ is the rotational velocity around the axis of symmetry ($V_{rot} = V_{obs} - V_c$ where $V_c$ is the velocity at the NH\textsubscript{3} integrated intensity peak). Following the methodology of \citet{Pineda2019TheFormation} we derive the specific angular momentum as a function of radius for CB 17 SMM1.\footnote{Using the 'velocity tools' package \url{https://github.com/jpinedaf/velocity_tools}} Figure \ref{fig:ang_moment} shows the resultant specific angular momentum profile for CB 17 and the comparison with the profiles derived for Class 0 sources in \citet{Pineda2019TheFormation} and \citet{Gaudel2020AngularEnvelopes}. 

\citet{Pineda2019TheFormation} studied two Class 0 and one FHSC candidate (L1451-mm). They found that the radial profiles of their three sources are consistent with the power-law $j(r) \propto r^{1.80\pm0.04}$ down to 1,000au.
\citet{Gaudel2020AngularEnvelopes} find that the specific angular momentum profiles of young Class 0 objects can be fit by a broken power-law of $j(r) \propto r^{0.3\pm{0.3}}$ inward of 1,600 au and $j(r) \propto r^{1.6\pm{0.2}}$ otherwise.

The power-law for the larger radii in \citet{Gaudel2020AngularEnvelopes} has a similar exponent to the \citet{Pineda2019TheFormation} power-law. Both studies conclude that the observed specific angular momentum at these scales is likely not the result of purely rotational motion, suggesting that gravitationally-driven turbulence from large-scale collapse motions or turbulent cascades from the large-scale molecular cloud could be the source of this outer envelope angular momentum.

We find that the specific angular momentum radial profile of CB 17 is in good agreement with the power-law relation from \citet{Pineda2019TheFormation} and the power-law relation from \citet{Gaudel2020AngularEnvelopes} at r > 1,600 au (see Figure \ref{fig:ang_moment}). It is also in good agreement with that of the star-forming filament LBS23, where $j(r) \propto r^{1.83\pm0.01}$ \citep{Hsieh2021RotatingFilament}. 


The close alignment between the CB 17 specific angular momentum profile and the power-law relations from \citet{Pineda2019TheFormation} and \citet{Gaudel2020AngularEnvelopes} is interesting given that the all of the sources used in fitting the power-laws in these published studies  already formed a compact object in their center and are thus more evolved than CB 17 SMM1. This implies that the average motions at 1,000 au scales are not greatly affected by the presence of a central object, and might instead originate from larger-scale turbulent cascades or 
colliding turbulent flows \citep{Chen2019InvestigatingGBT-Argus} as we suggested in Section~\ref{subsubsec:vel}. The former would be in agreement with the measurements in \cite{Hsieh2021RotatingFilament}, supporting the same mechanism producing velocity gradients in a prestellar core and Class 0 protostars is also present at filamentary scales. Furthermore, our results show clear agreement  in the magnitude of the specific angular momentum profiles between our starless core and the Class 0 sources. This may suggest that these motions are preserved throughout the pre-stellar stage and then later inherited by the more evolved (protostellar) cores.
Future measurements of the specific angular momentum profile as a function of radius for a sample of isolated prestellar cores ( such as L1544 \citep{Crapsi2007Observing1544}), would help elucidate whether neighboring protostars, as is the case of CB 17 IRS, have any influence in core motions at $\sim 10^{3}$ au scales.

\section{Summary and Conclusion} \label{sec:conclu}

The primary goal of this study was to determine the evolutionary state of the first core candidate CB 17 MMS and assess whether or not this source is a bonafide first core. We revealed the continuum emission using NOEMA 3 mm observations with a resolution of $\sim$625 au and the gas properties with VLA observations of NH\textsubscript{3}(1,1) and NH\textsubscript{3}(2,2) with a resolution of 1,120 au. Our conclusions can be summarized as follows:
\begin{enumerate}
\setlength{\itemsep}{1pt}
  \item 
  NOEMA 3mm observations towards the FHSC candidate CB 17 MMS do not detect any compact source at the location of the 1.3mm compact source reported in \citet{Chen2012SUBMILLIMETERCORE}, with a significance of at least 9$\sigma$. We argue that CO outflow emission originally thought to originate in the FHSC candidate outflow actually corresponds to outflow emission from a nearby protostar that is deflected by the dense ambient material.
  \item
  We detect compact NH$_3$ emission showing a peak column density of $1.5\times10^{15}$ cm\textsuperscript{-2}, an average temperature of 10.4 K and subsonic motions. The line-of-sight temperature profile shows a slight drop of 1 K in the inner 2,000 au. 
  \item 
  The core exhibits a velocity gradient aligned with the neighboring outflow axis, showing an arc-like structure in the p-v diagram and an off-center region with high velocity dispersion caused by two distinct velocity peaks. We believe that these features indicate CB 17 SMM1 has two overlapping velocity components that are compressed near the center of the core due to interaction with the CO outflow from the nearby protostar.
  \item 
  We fit the radial column density profile and found that the profile exhibits a flat inner region out to $r_0=1,800$ au and a central density of $6\times10^5$ cm$^{-3}$. An analysis of the density profile and virial mass of the core reveal it to be close to being gravitationally bound. This is consistent with our understanding of this starless source as being in an early evolutionary stage, potentially triggered by the  compression caused the outflow.
 \item The specific angular momentum radial profile of the CB 17 starless core aligns closely that observed by others in more evolved Class 0 sources. This similarity between sources at very different evolutionary stages suggest that the dominant outer envelope motions for these sources are remnants of large-scale dense cloud motions which are preserved through the early stages of star formation.
\end{enumerate}


The results of our study lead to the conclusion that CB 17 SMM1 is fully consistent with a marginally bound starless core allowing us to definitively rule out CB 17 MMS as a bona fide FHSC. A follow up study using the 3 mm molecular lines observed with NOEMA (including deuterated and non-deuterated species) will help us shed light on the inner physical and chemical properties of this newly confirmed starless core. 

Finally, we note that from the early list of eight FHSC candidates that were proposed based on SED, compact emission and/or outflow detection, which include: Cha-MM1, Per-bolo 58, L1448 IRS 2E, L1451-mm, CB 17 MMS, B1b-N and B1b-S, Per-bolo 45 \citep{2006BellocheEvolutionary,Enoch2010ACore,Chen2010L1448Core,Pineda2011THEVeLLO,Chen2012SUBMILLIMETERCORE,Pezzuto2012HerschelCloud,2012SchneeStarless}), about half have now been confirmed to be starless cores, while the other half are most likely very young Class 0 protostars. Only one, L1451-mm, remains a promising FHSC candidate \citep{Pineda2011THEVeLLO, Maureira2020ALMACandidates}.

\acknowledgments
MJM thanks Jan-Martin Winters for the assistance and support with the NOEMA data calibration. 
The authors thank Munan Gong and Kedron Silsbee for useful discussions that helped improving the discussion of the results. This work is based on observations carried out under project number S20AD [081-20] with the IRAM NOEMA Interferometer [30m telescope]. IRAM is supported by INSU/CNRS (France), MPG (Germany) and IGN (Spain). SHS thanks the Science, Technology, and Research Scholars (STARS) II Program for their support. 
This work was partially funded by the National Science Foundation award AST-1714710, awarded to HGA.  


\software{CASA (\citealt{McMullin2007CASAApplications}), GILDAS (\citealt{2005Pety}; \citealt{2013Gildas}), pyspeckit (\citealt{Ginsburg2011PySpecKit:Toolkit})}


\vspace{3.8cm}

\bibliography{sample63.bib}
\bibliographystyle{aasjournal.bst}



\appendix

\section{Additional figure}
\begin{figure}[h]
\gridline{\fig{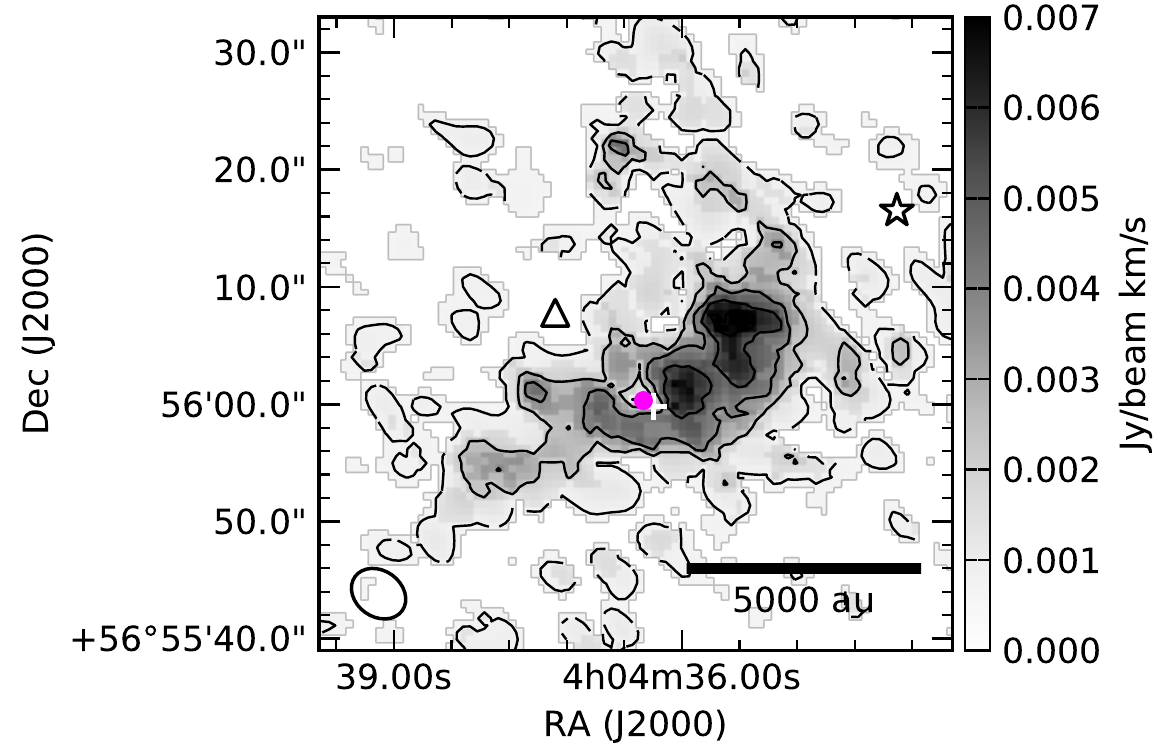}{0.5\textwidth}{}}
\caption{Integrated intensity map of NH$_3$(2,2). Black contours are drawn at 10\%, 30\%, 50\%, 70\%, and 90\% of the peak NH$_3$(2,2) integrated intensity. The peak NH\textsubscript{3} column density (as seen in Figure \ref{fig:mom0_t_ncol}) is marked by the white plus sign. The N\textsubscript{2}D+(3-2) peak emission from the SMA observations reported by \citet{Chen2012SUBMILLIMETERCORE} is indicated by the fuchsia circle. As we note in Section \ref{subsubsec:mass}, this emission peak is closely aligned with the NH\textsubscript{3} column density peak, but is not coincident with any NH\textsubscript{3}(2,2) emission peaks. The position of the Class I source CB 17 IRS  is marked in the upper right part of the panel with a star symbol. The region of high velocity dispersion, which can be seen in Figure \ref{fig:vel_sigma}, is marked by the $\Delta$ symbol. The black oval in the bottom left corner indicates the synthesized beam. 
\label{fig:mom0_22}}
\end{figure}


\end{document}